# Investigation of Dephasing In an Open Quantum System under Chaotic Influence via a Fractional Kohn-Sham Scheme

T. Ganesan

(E-mail: *tim.ganesan@gmail.com*)

**ABSTRACT**

In this work, the dynamics of dephasing (without relaxation) in the presence of a chaotic oscillator is theoretically investigated. The time-dependent density functional theory (TDDFT) framework was employed in tandem with the Lindblad master equation approach for modeling the dissipative dynamics. By employing the Kohn-Sham (K-S) scheme under certain approximations, the exact model for the potentials was acquired. In addition, a space-fractional K-S scheme was developed (using the modified Riemann-Liouville operator) for modeling the dephasing phenomenon. Extensive analyses and comparative studies were then done on the results obtained using the space-fractional K-S system and the conventional K-S system.

## Introduction

In most investigations involving computationally intense many-electron systems, time-dependent density functional theory (TDDFT) is the most commonly used approach. This is due to its practical utility, computational efficiency and accuracy [1-5]. TDDFT has been successfully employed to a vast amount of problems involving such systems to ascertain and understand their real-time dynamics. For instance, TDDFT has been applied in fields such as photochemical dynamics for the computation of the valence electron excitation energies in large chemical compounds [6,7]. Besides, TDDFT has also been employed to study characteristics of electronic interactions with cavity photons fields under the influence of quantized electromagnetic fields [8]. In addition, TDDFT has been implemented extensively in studies involving electron-phonon coupling in single molecular transport [9,10].

For modeling the dissipative dynamics of quantum systems, dissipative models such as: the Cladeira-Legett Model [11], Dissipative Two-level System [12], Dephasing Rate SP formula [13] and Ghirardi-Rimini-Weber (GRW) theory [14] are commonly employed. However, during the modeling of dissipative dynamics of a many-electron system, an alternate framework such as the master equation approach is commonly used in tandem with TDDFT. In this work, the open quantum system (OQS) is treated under the Markovian approximation and hence is suitable to be described by the Lindblad master equation [15, 16].

When solving interacting many-electron systems using theories like orbital-free density functional theory (DFT), the computations of the electronic interaction energies are often inaccurate and tedious. Thus, obtaining accurate approximations for the kinetic energy functional are difficult. A more accurate and computationally efficient technique is provided by the Kohn-Sham (K-S) scheme [17]. This approach works by the construction of an alternate non-interacting system with the same electronic density as the interacting system [18]. In Yuen-Zhou et al.(2009) [18] and Yuen-Zhou et

*T. Ganesan*
*Department of Chemical Engineering, Universiti Teknologi Petronas AQ8*
*E-mail: tim.ganesan@gmail.com*

al., (2010)[19] the existence of an OQS-TDDFT Kohn-Sham Scheme was proven by the Van Leeuwun construction[20] and the extension of the OQS Runge-Gross theorems to non-Markovian master equations were established. In those works, the closed and non-interacting K-S scheme was employed to generate the time-independent density of an interacting OQS. Similarly, this closed and non-interacting K-S scheme was employed in this work.

Most OQSs are commonly modeled in the context of a quantum well endowed with an external potential which acts as a simple harmonic oscillator[21-23]. In addition, dissipation or decoherence of OQSs are also usually studied by using a simple harmonic oscillator as an external potential [24, 25]. To the best of our knowledge, the effects of chaotic potential on the mechanism of dephasing in many-electron systems are rarely investigated. As can be seen in previous literature, some chaotic systems often present themselves in a complex or nonlinear form that is fractional in order [26, 27]. This is to say that many modeling efforts on such systems have identified that fractional-order calculus depicts the phenomena more accurately as compared to conventional approaches [28-30]. An example of such situations is the peroxidase—oxidase reactions which oscillates in a chaotic manner[31]. To model the interactions of such a reaction at a molecular level will mean to accurately model the quantum effects (such as the measurements process/decoherence) at those levels. Hence, the application of the fractional approach as proposed in this work may be targeted towards modeling the mentioned phenomena in such systems.

The first assimilation of fractional approaches into quantum mechanics was introduced by Laskin, (2008)[32]. In Laskin, (2008)[32], the quantum Reisz derivative[33] was introduced and efficiently applied to develop fractional concepts in physics such as the fractional Schrodinger equation and the fractional uncertainty principle. In Laskin, (2008)[32], the physical application of the newly developed fractional quantum mechanics was analyzed for a free particle system. Another work involving the fractionalization of the Schrodinger equation was done in Naber (2002)[34] and Rozmej and Bandrowski, (2010)[35]. In Naber (2002)[34], the Caputo fractional derivative[26] was used to fractionalize the time derivative to develop the time-fractional Schrodinger equation. The time-fractional and space-fractional Schrodinger equations were developed and analyzed in Rozmej and Bandrowski, (2010)[35] using the Riemann-Liouville (RL) fractional derivative and the Caputo fractional derivative (see Podlubny (1998)[26] for a more comprehensive discussion on fractional derivative operators). Another interesting work involving fractional quantum mechanics was done by Muslih et al., (2010)[36]. In that work the fractional Schrodinger equation was produced by using the fractional variation principle and a fractional Klein-Gordon equation. In Muslih et al., (2010)[36], the fractional Schrodinger equation was applied to ascertain the eigensolutions of a particle in an infinite potential well. In addition, fractional calculus has been utilized in quantum field theory[37] where concepts such as fractional phase space was constructed and utilized to develop field theories in fractal space. These concepts were used to identify ultraviolet cosmological solutions and their effects in the early universe. Fractional calculus has also been applied in particle physics[38, 39] where the fractional D'Alembertian operator was introduced and the fractional Lorentz transforms and fractional Klein-Gordon equations were developed. In addition this version of the Klein-Gordon equation has been employed for modeling linear dispersive phenomena[40]. Besides, nonlinear solution methods such as homotopy analysis have been implemented to solve the fractional Klein-Gordon equation[41].

The aim of this work is to study the pure dephasing phenomena (without relaxation) of an open interacting electronic system under the influence of a chaotic potential. By considering some suppositions, this system is transformed and solved as a closed non-interacting one-electron system. This is done by using TDDFT, and evolving it under the Lindblad master equation. By employing the K-S Scheme under the approximations established in Tempel and Aspuru-Guzik, (2011)[42], the exact model system for the potentials is acquired. Many of The assumptions and approximations employed in this work follow closely those in the work of Tempel and Aspuru-Guzik, (2011)[42]. This is due to the simplicity and efficiency of their exact potential formulation approach. Chaotic influence is then introduced into the one-electron system by using the delta-kicked harmonic oscillator[43] as an external potential. A space-fractional K-S scheme is developed using the modified RL operator[44]. This fractional scheme is then used for modeling the dephasing system driven by the chaotic potential. Detail analyses and comparative studies are then carried out on the exact potentials obtained using the space-fractional K-S system and the conventional or regular K-S system.

This article is organized as follows: Open Quantum Systems and Lindblad Master Equation sections provide a brief introduction on OQSs and the Lindblad master equation, respectively. Construction of Exact Potential for the Kohn–Sham Equations section presents the procedures involved in the construction of the exact potential for the regular K–S equations while Construction of Exact Potential for the Space-Fractional Kohn–Sham Equations section provides some details regarding the development of the space-fractional K–S equations. The Delta-Kicked Harmonic Oscillator section discusses and compares the results produced using the K–S and the space-fractional K–S equations during dephasing under the influence of the chaotic potential. Finally, this article ends with some concluding remarks and suggestions for future works.

## Open Quantum Systems

Open quantum systems are usually defined in the context of larger combined systems called the system and environment (*S+E*). Here the environment is another larger quantum system which is interacting with the quantum system, *S*. Therefore the state of the system, *S* would evolve non-unitarily according to its own internal dynamics and to the environment, *E*. Therefore, the reduced density operator is used to trace over the environment degrees of freedom to obtain the system's state as follows:

$$\rho_S = Tr_E\{\rho\} \qquad (1)$$

where $\rho_S$ is the system density and $\rho$ is the combined (*S+E*) density operator. The reduced system dynamics can then be described by obtaining the exact quantum master equation:

$$\frac{d\rho_S}{dt} = -i[H_s(t), \rho_S(t)] + \int_0^t d\tau \Delta(t-\tau)\rho_S(\tau) + \psi(t) \qquad (2)$$

where $\Delta(t-\tau)$ is the memory kernel, $\psi(t)$ is the inhomogeneous term from the initial interaction relations between the system and the environment and [.] denotes the commutator. Following the TDDFT, the electronic density, $n(r,t)$ and the number density operator, *N(r)* are as follows:

$$n(r,t) = Tr_S\{\rho_S(t) \cdot N(r)\} \qquad (3)$$

$$N(r) = \sum_{i=1}^{N} \delta(r-r_i) \qquad (4)$$

where $\delta(r-r_i)$ is the Dirac delta function. An auxiliary master equation evolving under a different Hamiltonian, $H'_S(t)$ can be constructed using the exact quantum master equation given in equation (2):

$$H'_S(t) = \frac{1}{2}\sum_{i=1}^{N}\nabla_i^2 + \sum_{i<j}^{N}\frac{\alpha}{|r_i-r_j|} + \sum_{i=1}^{N}V'(\alpha,r_i,t) \qquad (5)$$

$$\frac{d\rho'_S}{dt} = -i[H'_s(t), \rho'_S(t)] \\ + \int_0^t d\tau \Delta'(t-\tau)\rho'_S(\tau) + \psi'(t) \quad (6)$$

where $\alpha$ is the interaction strength while $\Delta'(t-\tau)$ and $\psi'(t)$ are the auxiliary memory kernel and the inhomogeneous term respectively. Following the Van Leeuwun construction[20], the existence of a unique local potential, $V'(\alpha,r,t)$ for the auxiliary system at any $\alpha$ and arbitrary $\Delta'(t-\tau)$ and $\psi(t)$ can be proven[19]. The central idea presented in Yuen-Zhou et al., (2010)[19] is that for any open interacting system, an auxiliary system can be constructed such that the potential $V'(\alpha,r,t)$ imposes the correct density evolution with a different set of electron-electron and system-bath interactions. In Tempel et al., (2011)[42], the auxiliary open but non-interacting Kohn-Sham system is obtained by setting $\alpha = 0$ while maintaining and $\psi'(t) = \psi(t)$. The closed and non-interacting Kohn-Sham system is obtained by setting $\alpha = 0$, $\Delta'(t-\tau) = 0$ and $\psi'(t) = 0$.

## Lindblad Master Equation

Due to the difficulty in solving the full system-bath evolution dynamics, the Markov Approximation where the memory kernel is in local time is imposed. The memory kernel is represented as follows:

$$\hat{D}\rho'_S(t) = \int_0^t d\tau \Delta'(t-\tau)\rho'_S(\tau) \quad (7)$$

The Markov approximation is only effective if the system relaxation time (to thermal equilibrium) is greater than maximal correlation time of the bath. The Lindblad master equation is as follows:

$$\hat{D}\rho'_S(t) = \sum_{i=1} 2L_i \rho(t) L_i^+ - L_i^+ L_i \rho(t) \\ - \rho(t) L_i^+ L_i \quad (8)$$

The Lindblad operators $L_i$ and $L_i^+$ describes transitions between eigenstates of the system Hamiltonian, $H_s$ during the electronic interaction with the environment. In this work, a two-level system undergoing pure dephasing without relaxation is considered. Therefore, this is a completely elastic and adiabatic process with no energy transfer between the system and the environment. The Lindblad operators and the system Hamiltonian, $H_s$ are as follows:

$$L_1 = L_1^+ = \sqrt{\gamma}\begin{pmatrix} 1 & 0 \\ 0 & 0 \end{pmatrix} \quad (9)$$

$$H_S = \begin{pmatrix} E_0 & 0 \\ 0 & E_1 \end{pmatrix} \quad (10)$$

where $\gamma$ is the rate of state collapse due to dephasing while $E_0$ and $E_1$ are the system's initial energy and final energy states. Therefore the taking into account the Lindblad operators, the master equation is solved to obtain a 2x2 matrix with the entries $\rho_{11} = \rho_{11}(0)$, $\rho_{12} = \rho_{12}(0)e^{-\gamma t - i(E_1-E_0)}$ $\rho_{21} = \rho_{21}(0)e^{-\gamma t - i(E_1-E_0)}$ and $\rho_{22} = \rho_{22}(0)$. Thus, the system's non-unitary evolution during pure dephasing is described by the solution to the master equation.

## Construction of Exact Potential for the Kohn-Sham Equations

The construction of the exact potential for the K-S equations has been carried out successfully in Tempel et al., (2011)[42]. These procedures are reproduced in this work and are briefly outlined in this section. By solving the Lindblad master equation exactly, the density evolution of the one-electron system, $\rho(t)$ can be obtained. Throughout this paper, the formulations presented would be limited to one spatial dimension for the sake of simplicity. However, these formulations could readily be implemented for higher spatial dimensions. For the sake of compactness in the representation of the formulations, the following notations are used, $\partial f/\partial x = \partial_x f$ and $\partial^2 f/\partial x^2 = \partial_{xx} f$. The evolving K-S equation is given as follows:

$$i\partial_t \phi = V_{KS}\phi - \frac{1}{2}\partial_{xx}\phi \tag{11}$$

such that the densities are represented as:

$$n(x,t) = |\phi(x,t)|^2 = Tr_s\{\rho_s(t)N(x)\} \tag{12}$$

For a one-electron system, the exchange potential cancels off the Hartree potential. Thus, the K-S potential, $V_{KS}$ is reduced to the external potential, $V_{ext}$ and the OQS correlation potential, $V_c$:

$$V_{KS} = V_{ext} + V_c \tag{13}$$

First, to obtain the exact correlation potential, the ansatz which is the K-S orbital for a single electron is defined as the following:

$$\phi(x,t) = \sqrt{n(x,t)}\exp(i\theta(x,t)) \tag{14}$$

where $\theta(x,t)$ is the phase angle. Since the K-S density, $\phi(x,t) = f(n(x,t),\theta(x,t))$, therefore the following chain rules can be employed to compute the exact correlation potential:

$$\partial_t \phi = \partial_\theta \phi \cdot \partial_t \theta + \partial_n \phi \cdot \partial_t n \tag{15}$$

$$\partial_{xx}\phi = \partial_n\phi \cdot \partial_{xx}n + \partial_\theta\phi \cdot \partial_{xx}\theta + \partial_{nn}\phi \cdot (\partial_x n)^2 + 2\partial_{n\theta}\phi \cdot \partial_x\theta \cdot \partial_x n \tag{16}$$
$$+ \partial_{\theta\theta}\phi \cdot (\partial_x\theta)^2$$

where for simplicity $n(x,t)$ and $\theta(x,t)$ are denoted as $n$ and $\theta$. The exact K-S potential obtained in terms of $n$ and $\theta$ are as follows:

$$V_c = -\partial_t\theta + \frac{1}{4n}\partial_{xx}n - \frac{1}{8n^2}(\partial_x n)^2$$
$$-\frac{1}{2}(\partial_x\theta)^2 - V_{ext} \tag{17}$$

## Construction of Exact Potential for the Space-Fractional Kohn-Sham Equations

In this work, the K-S equation is represented in a fractional form. The fractional derivative used here is the modified RL operator developed in Jumarie (2006)[44]. This operator has proven to have been very useful and simple to be implemented for modeling nonlinear and complex phenomena[45, 46]. As can be seen in equation (17) the spatial derivatives are the nonlinear terms that heavily influence the dynamics of the exact correlation potential and in effect the K-S potential. Thus similar to Laskin, (2002)[33] and Rozmej and Bandrowski, (2010)[47], the spatial component of the K-S equation is fractionalized in this work to capture these nonlinearities. The space-fractional K-S equation is defined as follows:

$$i\partial_t\phi = V_{KS}\phi - \frac{1}{2}\partial_x^\alpha\phi \tag{18}$$

where $\partial_x^\alpha$ is the spatial derivative using the modified RL operator[44] and $\alpha \in (0,1)$ is the fractional variable. The fractional K-S orbital is defined as follows:

$$\phi = \sqrt{n}E_\alpha(i\theta) \tag{19}$$

where $E_\alpha(\cdot)$ is the Mittag-Leffler function[48] which is the fractional analogue of exponentiation. In fractional calculus, three types of chain rules may be defined due to the following property[49]:

$$\partial_x^\alpha \left(x^{2\alpha}\right) \neq \partial_x^\alpha \left(x^2\right)^\alpha \neq \partial_x^\alpha \left(x^\alpha\right)^2 \quad (20)$$

Hence, one of the three chain rules is used in this work. The chain rule is chosen due to the properties of the functional $\phi$ and the functions $n$ and $\theta$. First, the functional $\phi$ in equation (19) is $\alpha$-differentiable. Second $n$ and $\theta$ are differentiable in the usual way. The selected chain rule is shown by equation A3 in the appendix: Following the chain rule the spatial derivative in equation (18) is formulated:

$$\partial_x^\alpha \phi = \partial_n^\alpha \phi \cdot (\partial_x n)^\alpha + \partial_\theta^\alpha \phi \cdot (\partial_x \theta)^\alpha \quad (21)$$

The electron density and phase angle derivative terms in equation (22) are computed using the Leibniz product law and the differential property given in equations A1 and A5 in the Appendix). By applying the properties given in equations A8 and A9, the following K-S density derivatives are obtained:

$$\partial_n^\alpha \phi = E_\alpha(i\theta)\Gamma\left(\frac{3}{2}\right)\Gamma^{-1}\left(\frac{3}{2} - \alpha\right) n^{\frac{1}{2} - \alpha} \quad (22)$$

$$\partial_\theta^\alpha \phi = i\sqrt{n}\alpha^{-\alpha}\theta^{1-\alpha} E_\alpha(i\theta) \quad (23)$$

where $\Gamma(\cdot)$ is the gamma function. Therefore, the spatial derivative of the K-S equation is as follows:

$$\partial_x^\alpha \phi = \left[E_\alpha(i\theta)\Gamma\left(\frac{3}{2}\right)\Gamma^{-1}\left(\frac{3}{2} - \alpha\right)\right] \times \\ n^{\frac{1}{2} - \alpha} \partial_x^\alpha n + \left[i\sqrt{n}\alpha^{-\alpha}\theta^{1-\alpha} E_\alpha(i\theta)\right]\partial_x^\alpha \theta \quad (24)$$

The Mittag-Leffler series (equation A2) is then truncated at the second order term. Therefore, the fractional K-S orbital becomes:

$$\phi = \sqrt{n} + \frac{i\theta\sqrt{n}}{\Gamma(1+\alpha)} \quad (25)$$

By using the fractional K-S orbital as the ansatz and proceeding with some manipulations, the fractional correlation potential is obtained:

$$\tilde{V}_c = -\left(\frac{\partial\theta}{\partial t}\right)\left(\frac{\Gamma(1+\alpha)}{\Gamma(1+\alpha)^2 + \theta^2}\right) \\ + \left[\frac{1}{2\sqrt{n}}\Gamma\left(\frac{3}{2}\right)\Gamma^{-1}\left(\frac{3}{2} - \alpha\right) n^{\frac{1}{2} - \alpha} \times \right. \quad (26) \\ \left.\left(\frac{\partial n}{\partial x}\right)^\alpha\right] - V_{ext}$$

**Delta-Kicked Harmonic Oscillator**

The delta-kicked harmonic oscillator was developed in Zaslavsky et al., (1991)[50] for modeling the motion of a charged particle in a homogeneous static magnetic field with a orthogonal propagating electric field wave packet. The dynamical behavior of the one-dimensional delta-kicked oscillator has been studied extensively[43, 51, 52]. The delta-kicked harmonic oscillator has also been employed for modeling electron transport in semiconductor super-lattices[53, 54] and for modeling atomic optics using ion traps[55]. One property of the delta-kicked harmonic oscillator which varies from the harmonic oscillator is the frequency. Unlike the harmonic oscillator, the delta-kicked harmonic oscillator operates using two frequencies. Similar to its counterpart, the delta-kicked harmonic oscillator has a natural frequency related to its free oscillation. However, the delta-kicked harmonic oscillator also has another frequency associated to the periodic kicking of the potential. The delta-kicked harmonic potential is given as follows:

$$V_\delta(x,t) = -\left(\frac{m\omega^2 x^2}{2}\right) \\ + K\cos(kx)\sum_{-\infty}^{+\infty} \delta(t - n\tau) \quad (27)$$

where $m$ is the particle mass, $\omega$ is the harmonic frequency, $k$ is the kick wavenumber, $K$ is the kick strength, $\tau$ is the time interval between kicks and $\delta(\cdot)$ is the Dirac delta function. In this work, the delta-kicked harmonic potential, $V_\delta(x,t)$ was used as an external potential during the dephasing of the one-particle system. The dynamical behavior of the system is then analyzed using the TDDFT-OQS in tandem with the K-S equation and the Space-fractional K-S equation in the next section.

## Results & Analysis

In this work, the behavior and evolution of the approximated potentials and the electron densities are obtained using the regular and the space-fractional K-S schemes. A comparative analysis of the one-electron system under pure dephasing without relaxation is done for two cases. The first case is when the one-electron system is under the influence of an external potential (which is a harmonic oscillator) while the second exposes the system to a chaotic external potential (delta-kicked harmonic oscillator). The initial parameters used in this work for the system evolution are given in Table 1:

| Table 1. System Parameters and Physical Constants | |
|---|---|
| Parameters | Values |
| Lindblad parameters; $\gamma_0 = \gamma_1$ | 0.15 a.u. |
| Initial Density $\rho$ | (0.5, 0.5, 0.5, 0.5) |
| Ground state wave length, $\lambda_e$ | 5.5 x 10$^{-5}$ m |

Throughout this work the Planck constant, $\hbar$, the speed of light, $c$ and the electron mass, $m_e$ are set to one. The initial state of the OQS is set as a two-level system, $|\psi\rangle = 1/\sqrt{2}(|0\rangle + |1\rangle)$. The dephasing timescale, $\tau_{dec}^{mn}$ can be computed as follows:

$$\tau_{dec}^{mn} = 0.5(\gamma_m + \gamma_n) \qquad (28)$$

where $m$ and $n$ are the quantum states where in this work is two-level, $|0\rangle$ and $|1\rangle$. Therefore the dephasing timescale obtained is 0.15 a.u. The Lindblad parameters are selected accordingly to impose a weak system-environment coupling so that the dephasing takes place at a lengthy period of time.

It was observed that the space-fractional K-S and correlation potential for all fractional coefficient, $\alpha$ except for $\alpha = 0.3$ and $\alpha = 0.7$ contain singularities when $x \leq 0$. This situation recurs throughout the system's evolution with any type of external potential. Therefore, throughout this work the space-fractional potentials are confined to $\alpha = 0.3$. This behavior is due to the nature of the space-fractional potential functional that inhibit modeling for various fractional coefficients.

Besides, it was found that the obtained fractional K-S and correlation potentials obtained at $\alpha = 0.3$ contain a singularity at x = 0. Thus, in the subsequent analysis of the maps, this singularity is averaged over its neighboring points for functional regularity. It is highly possible that the fractional K-S potential does not exist at these ranges. The evolution of the electron density, *n(x,t)*, the regular and the space-fractional K-S potentials are obtained using the formulations discussed in the previous sections. The evolution and spatial distribution of the K-S potential does not depend on the external potential. Nevertheless, some interesting aspects of the fractional K-S potentials can be observed when compared with the regular K-S potential in the presence of the delta-kicked harmonic oscillator. The time-dependent behavior of these potentials with the electron density as a reference is presented in Figure 1.

The dephasing in Figure 1 is represented clearly by the changing of the electron density distribution with respect to time. Therefore the density of the one-electron system gradually evolves into a mixed state (equilibrium) where $\rho_{00} = \rho_{11} = 0.5$ and $\rho_{01} = \rho_{10} = 0$. The mixed state is the total sum of the electronic ground state and its excited states. In Figure 1 it can also be observed that the regular and the space-fractional K-S potential are symmetric throughout and decay with time. When the electron density is symmetric, the coherent contribution from the population is maximum.

On the other hand, of it is asymmetric then the coherent contribution from the density is at maximum. Therefore, in the evolution shown in Figure 1, coherent contribution from the population is at maximum.

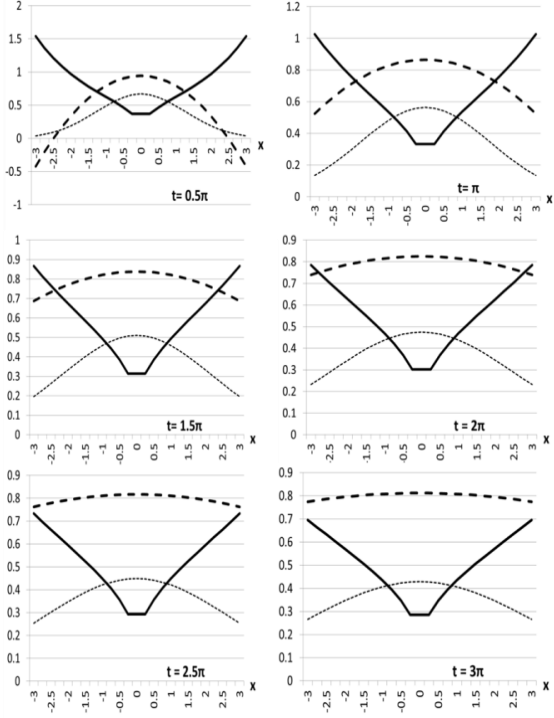

Figure 1. The spatial distribution and time evolution of the electron density, *n(x,t)* (dotted), regular K-S potential, $V_{KS}$ (dashed) and space-fractional K-S potential, $\widetilde{V}_{KS}$ (solid) where the external potential is a delta-kicked harmonic oscillator.

Besides, the space-fractional K-S potential is seen in Figure 1 to be increasingly attractive at the extremes of *x* generating a *U* - shaped curve which is a direct contrast to the regular K-S potential. The attractiveness of the regular and space-fractional K-S potential decays as time increases. This behavior is due to the decay of the correlation potential which directly affects the K-S potential (see equation (13)). Thus, as this happens the system becomes fully dominated by the external chaotic potential which results in increased repulsiveness and decreased attractiveness.

The other significant difference between the regular and space-fractional K-S potential is the way the distributions stabilize with time such that they gradually lose their intersection points. In addition, at time $t = 0.5\pi$ the regular K-S potential is repulsive at the spatial extremes (*x* < -0.5 and *x* > 0.5) while the space-fractional K-S potential is never repulsive throughout its evolution.

The space-fractional correlation potential, $\widetilde{V}_c$ was computed using the space fractional K-S equation. Figure 2 shows the profiles of the space-fractional correlation potential, $\widetilde{V}_c$ with respect to the regular correlation potential, $V_c$ at different times:

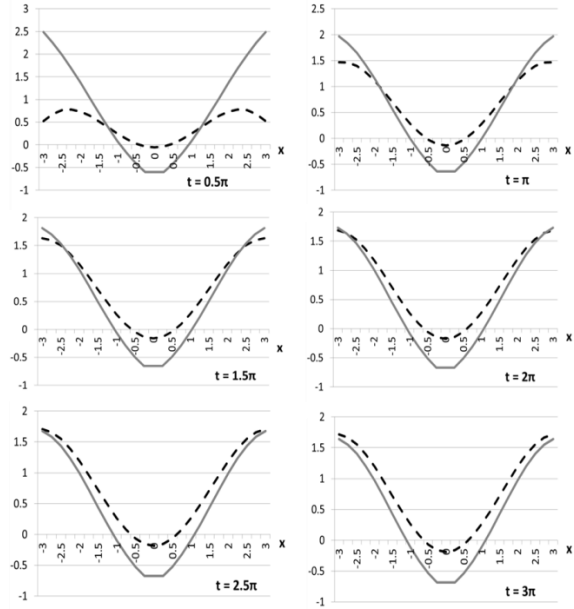

Figure 2: The spatial distribution and time evolution of the regular correlation potential, $V_c$ (dashed) and space-fractional correlation potential, $\widetilde{V}_c$ (solid) where the external potential is a delta-kicked harmonic oscillator

The correlation profiles shown in Figure 2 were taken at $\alpha = 0.3$. The delta-kicked oscillator was at the angular frequency, $\omega = 0.1$, kick strength, *K* = 1, kick wave number, *k* = 1 and the time interval between kicks, $\tau = 0.1$. It can be observed in Figure 2 that as time increases, with -1 < x < 1, the regular correlation potential gradually becomes attractive after $t = 0.5\pi$ getting closer to the values of the space-fractional correlation potential. This behavior may be due to the reduction in coherent contribution from the population as time increases resulting in a more attractive correlation potential. However, from Figure 2, it can be seen that this effect almost does not occur at all to the space-fractional correlation potential although there is a slight reduction to the attractiveness of the space-fractional correlation potential and a certain amount of repulsion during its evolution. According to the results given the space-fractional potential, it is possible that the reduction in coherent contribution from the population is not as significant as depicted using the regular correlation potential. In Figure 3, the profiles of the space-fractional correlation potential, $\widetilde{V}_c$ and the regular correlation potential, $V_c$ at various angular frequencies, $\omega$ of the delta-kicked potential is given:

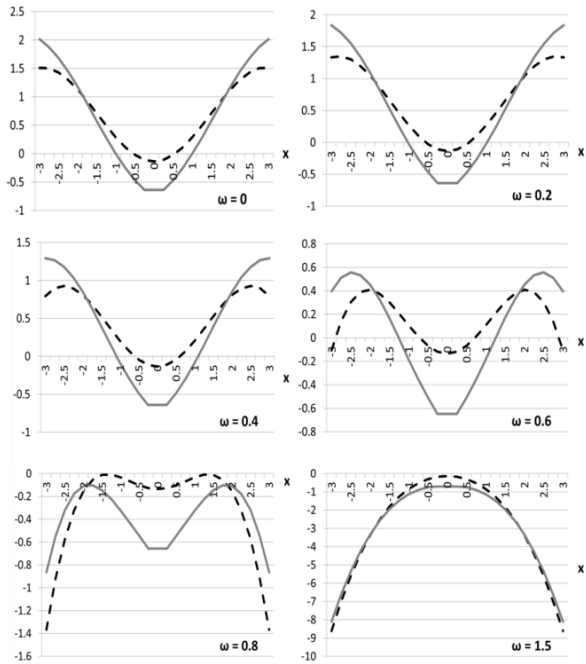

Figure 3: The spatial distribution and time evolution of the regular correlation potential, $V_c$ (dashed) and space-fractional correlation potential, $\widetilde{V}_c$ (solid) where the external potential is a delta-kicked harmonic oscillator with varied angular frequency, $\omega$

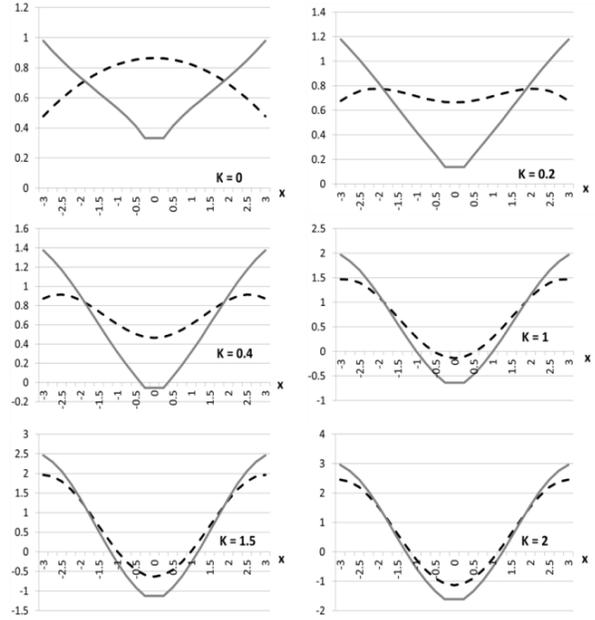

Figure 4: The spatial distribution and time evolution of the regular correlation potential, $V_c$ (dashed) and space-fractional correlation potential, $\widetilde{V}_c$ (solid) where the external potential is a delta-kicked harmonic oscillator with varied kick strength, $K$

The correlation potential distributions given in Figure 3 were taken at $\alpha = 0.3$ and at time, $t = \pi$. The delta-kicked oscillator was at the kick strength, $K$ = 1, kick wave number, $k$ = 1 and the time interval between kicks, $\tau = 0.1$. Similar to Figure 2, the space-fractional correlation potential distribution is more repulsive as compared to the regular correlation potential. Besides, at angular frequencies $0 \leq \omega \leq 0.6$, of the delta-kicked oscillator, the correlation potential is more attractive relative to the regular correlation potential at the extremes of x.

Both correlation potentials are seen to form a double–well hill structure at $0.4 \leq \omega \leq 0.8$. This is the intermediate state of equilibrium (at $w = 0.6$), where the system is driven into equilibrium by the external potential before its dynamics are fully dominated by the external potential. It is very interesting to observe that the space-fractional correlation potential forms a more prominent double–well hill structure at $w = 0.6$ and $w = 0.8$. Although it vaguely resembles the form of the regular correlation potential, the space-fractional correlation potential shown to be more repulsive at these regions. This may be some indication, that the fractional approach may shed some light on the dissipation dynamics of equilibrium structures that emerge and disperse when systems are gradually exposed to chaotic influence.

Both potentials become repulsive as the dynamics of the one-electron system is fully influenced by the high angular frequency, $\omega$ of the delta-kicked oscillator hence becoming equivalent, $\widetilde{V}_c = V_c$ at approximately $\omega >$ 1.5. In these instances the dynamics of the system is fully influenced by the high angular frequency of the delta-kicked oscillator. Figure 4 portrays the spatial distributions of the space-fractional correlation potential, $\widetilde{V}_c$ and the regular correlation potential, $V_c$ at kick strengths, $K$ of the delta-kicked potential is given.

The correlation potential distributions given in Figure 4 were taken at $\alpha = 0.3$ and at time, $t = \pi$. The delta-kicked oscillator was at the angular frequency, $\omega = 0.1$ where the kick wave number, $k = 1$ and the time interval between kicks, $\tau = 0.1$. The space-fractional correlation potential distribution in Figure 4 is observed to become increasingly repulsive at $K > 0.2$. At $K \geq 0.2$ both potentials are attractive. As the kick strength increases, the attractiveness of both correlation potentials decreases while the repulsiveness of the potentials increase. This trend progresses until the space-fractional correlation and the regular correlation potential profiles become identical (which happens at high oscillator kick strengths). Similar to the findings in Kells (2005)[51] at large kick strength it is possible that the stationary states of the populations undergo extensions. Due to these extensions, it can be seen that at $K \geq 1.5$, not much changes occur in the distributions of the regular and space-fractional fractional correlation potentials.

**Concluding Remarks & Central Findings**

In this work, the dephasing phenomenon of a one-particle system was modeled using the conventional K-S and space-fractional K-S systems in the context of the master equation approach. Information on the dynamics of dephasing is crucial for our understanding of the quantum measurement process. As can be seen in previous works, in the face of chaos and some forms of nonlinearities, the fractional formulations tend to depict and explain the system dynamics more accurately than conventional methods. It can be seen that in this work, that there exist critical differences between these two approaches when modeling dephasing phenomenon under chaotic influence. The most interesting results are observed to happen during the transition states (right before the external potential dominates) where the system reaches equilibrium (Figures 3 and 4). In addition, the evolution of the interplay between the attractiveness and the repulsiveness of the space-fractional correlation potential also offers some interesting insights (Figure 2).

In future works, the space-fractional K-S system and the K-S system under chaotic influence could be tested against experimental data to deepen our understanding with regards to dephasing phenomenon of such systems. Besides, other properties such as dissipative and Hamiltonian currents could also be studied in the context of a fractional framework. These studies could be extended to non-adiabatic systems undergoing relaxation per se as well as relaxation and dephasing simultaneously (complete decoherence). In this work, for simplicity the bath was represented using the Lindblad operators which were dependent only on the dephasing rate scalar. In the future, coupling position and momentum to the bath would be a viable method to study the system dynamics more accurately. Besides, the fractional approach could be to many-electron systems by analyzing the density evolution through the advancement of the Fock state occupation. This extension would shed some light on the physics of such systems.

Another interesting aspect is the existence of a Van Leeuwen Theorem for the fractional Kohn-Sham scheme. Even if such a direct mapping is not possible, perhaps future investigations could be channeled to explore if there exist some alternate mapping technique. Nevertheless even if a Van Leeuwen Theorem cannot be proved, this would provide useful insights on the structure of the K-S and space-fractional K-S schemes.

## Acknowledgments

*The author is very grateful to the reviewers for their efforts in providing useful and constructive insights that has been overlooked by the author during the preparation of the manuscript. The author would also like to thank the Chemical Engineering Department of Universiti Teknologi Petronas for their support during this research work.*

**Keywords:** dephasing, space-fractional Kohn-Sham scheme, exact model, density functional theory, Lindblad master equation.

# Appendix A: Fractional Calculus Formulae

1. Leibniz Product Law

$$\partial_x^\alpha (u \cdot v) = v \cdot \partial_x^\alpha u + u \cdot \partial_x^\alpha v \quad (A1)$$

2. Mittag-Leffler Expansion

$$E_\alpha(u) = \sum_{k=0}^\infty \frac{u^k}{(\alpha k)!} \quad (A2)$$

3. Chain Rule

$$\partial_x^\alpha f = \partial_u^\alpha f \cdot (\partial_x u)^\alpha \quad (A3)$$

4. Differentiation Properties

- If $c \in \Re$, then $\partial_x^\alpha c = 0$ \quad (A4)
- If $c \in \Re$, $\partial_x^\alpha [cf(x)] = c \partial_x [f(x)]$ \quad (A5)
- $\partial_x^\alpha [f(x(t))] = \partial_x [f(x)] \cdot x^\alpha(t)$ \quad (A6)
- $\partial_x^\alpha [f(x)] = \frac{1}{\Gamma(-\alpha)} \int_0^x (x-t)^{-\alpha} (f(t) - f(0)) dt$ \quad (A7)
- $\partial_x^\alpha [E_\alpha(\lambda x)] = \lambda \alpha^{-\alpha} x^{1-\alpha} E(\lambda x)$, $\lambda \in \Re$ \quad (A8)
- $\partial_x^\alpha [x^\gamma] = \Gamma(\gamma+1) \cdot \Gamma^{-1}(\gamma+1-\alpha) x^{\gamma-\alpha}$ for $\gamma > 0$ \quad (A9)

**GRAPHICAL ABSTRACT**

AUTHOR NAMES

T. Ganesan

TITLE

Investigation of Dephasing In an Open Quantum System under Chaotic Influence via a Fractional Kohn-Sham Scheme

TEXT

The dephasing of interacting many-electron systems under chaotic influence is an unexplored complex phenomenon. An effort towards a more complete theoretical understanding of its dynamics is carried out by employing fractional formulations. Using this approach some interesting insights are uncovered and these findings are examined in detail.

GRAPHICAL ABSTRACT FIGURE

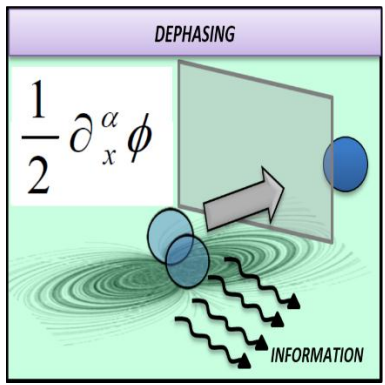